# Miniaturized microscopes to study neural dynamics in freely-behaving animals

Weijian Zong[1,*], Weijian Yang[2,#]
[1]Kavli Institute for Systems Neuroscience and Centre for Algorithms in the Cortex, Norwegian University of Science and Technology (NTNU), Trondheim NO-7491, Norway
[2]Deparatment of Electrical and Computer Engineering, University of California, Davis, Davis, CA 95616, USA
* weijian.zong@ntnu.no
# wejyang@ucdavis.edu

Abstract: Head-mounted miniaturized microscopes, commonly known as miniscopes, have undergone rapid development and seen widespread adoption over the past two decades, enabling the imaging of neural activity in freely-behaving animals such as rodents, songbirds, and non-human primates. These miniscopes facilitate numerous studies that are not feasible with head-fixed preparations. Recent advancements have enhanced their capabilities, allowing for faster imaging, larger fields of view, and deeper brain penetration. In this review, we examine the latest progress in one-photon and multi-photon miniscopes. We highlight the unique opportunities these devices present for neuroscience research, discuss the current technical challenges, and explore emerging technologies that promise to advance the development of miniscopes.

---

One fundamental question in neuroscience is how animal behavior and perception emerge from the activity of neural circuits, which are often composed of hundreds of thousands of interconnected neurons. Optical methods, coupled with activity indicators, are ideally suited to study this question as they can image large populations of neurons with specific cell-types in fine, often subcellular, resolution, over days and weeks. Benchtop microscopes with head-fixed preparations of animals are the most mature approach. Though they can study a wide range of behaviors using virtual reality[1,2], there are challenges in studying the neural circuits under specific behaviors such as 2D or 3D spatial navigation, social behavior, fear and escape responses, and sleep. Head-fixed or body-fixed conditions may alter neural representations compared to those in nature states[3-6]. Miniaturized microscopes (also known as miniscopes), which can be mounted on the animal's head to image its neural activity while it freely behaves, offer a new solution. Pioneered by Tank in the early 2000s with a two-photon version[7], miniscopes have flourished over the past two decades[8-39], in particular with the one-photon (1P) versions gaining wide adoption[9,15]. Recently, there has been a resurgence of interest in two-photon (2P) miniscopes[29-36], along with the introduction of three-photon (3P) versions[37-39]. In this review, we examine the architecture and state-of-the-art of miniscopes and provide readers guidance on selecting specific types. We discuss the challenges faced by these miniscopes and explore emerging techniques, offering a roadmap for future development.

## Why miniscopes

Miniscopes are uniquely suited in studying cell-type specific neuronal activity across a large population of neurons during animal behaviors that are challenging or impossible to observe with head-fixed preparations. One important question is how the animals navigate in space. In rodents, while virtual reality in head-fixed setups can simulate navigation tasks, it is typically 1D[1,2]. For 2D navigation, the lack of head-rotation or body-rotation (but see Ref.[40,41]) limits the vestibular input and disrupts eye-head movement coupling, raising concerns about whether the neuronal circuits observed in head-fixed or body-fixed condition accurately reflect natural states[3-5]. For 3D navigation tasks, the head-fixed setups are entirely inadequate. Other behaviors that are better suited to study through miniscopes include social interactions, foraging and hunting, fear and escape response, and sleep (Fig. 1). In non-human primates, neurons were found to encode the same actions differently depending on the context: activity patterns were more complex in freely moving conditions compared to when the body was restrained[6]. In songbirds, training them to sing under head-fixed conditions is extremely challenging[42], and the neural coding of song in such a state may differ due to altered motivational or behavioral contexts compared to that in freely moving condition. This highlights the need for head-mounted devices to study natural singing behavior. While electrical recording, typically through electrode arrays, is also a powerful tool for recording neural activity in freely-moving animals, the number of cells they can record in a local region is typically much smaller than that in miniscopes. Furthermore, it may be challenging to distinguish the cell-type and the spatial locations of the recorded cells, recording dynamics from subcellular structures such as dendrites and spines, as well as conduct longitudinal study with electrode arrays.

The past two decades have seen widespread adoption of miniscopes, primarily one-photon, across various species, behavioral paradigms, brain regions, cell types and structures (Fig. 1) to study the dynamics of the neural system where it is challenging with benchtop microscopy or electrophysiology. Many studies focus on imaging deep brain regions in mice, such as the hippocampus[10,12,43], thalamus[44], the anterior cingulate cortex[45,46], hypothalamus[47,48] and brain stem[49] as these areas are responsible for navigation, memory, social behavior, sleep and motion—behavioral paradigms that are not easily studied in head-fixed conditions. There are also studies on songbirds[19,50], bats[51] and non-human primates[52] (Fig. 1). Recently, 2P miniscopes have gained substantial interest due to its high spatial resolution and imaging quality in scattering tissue, enabling imaging of fine features such as dendrites and spines[29,53]. By leveraging the various fluorescent indicators, miniscopes have been employed to investigate a wide range of biological structures, including spines and dendrits[25,29,31,53], neurons[8-13,15,16,18-22,27,29,30,32-39], astrocytes[36], microglia[54] and blood vessels[9,23] (Fig. 1), as well as diverse physiological signals such as membrane potential[55], calcium dynamics[8-22,25,27,29-39], neurotransmitter activity[56], and blood flow[9]. Different designs of miniscopes have been adapted for population imaging across various spatial scales in the neural system, ranging from multiple regions across the brain, to local circuitry, and to subcellular structures (Fig. 1).

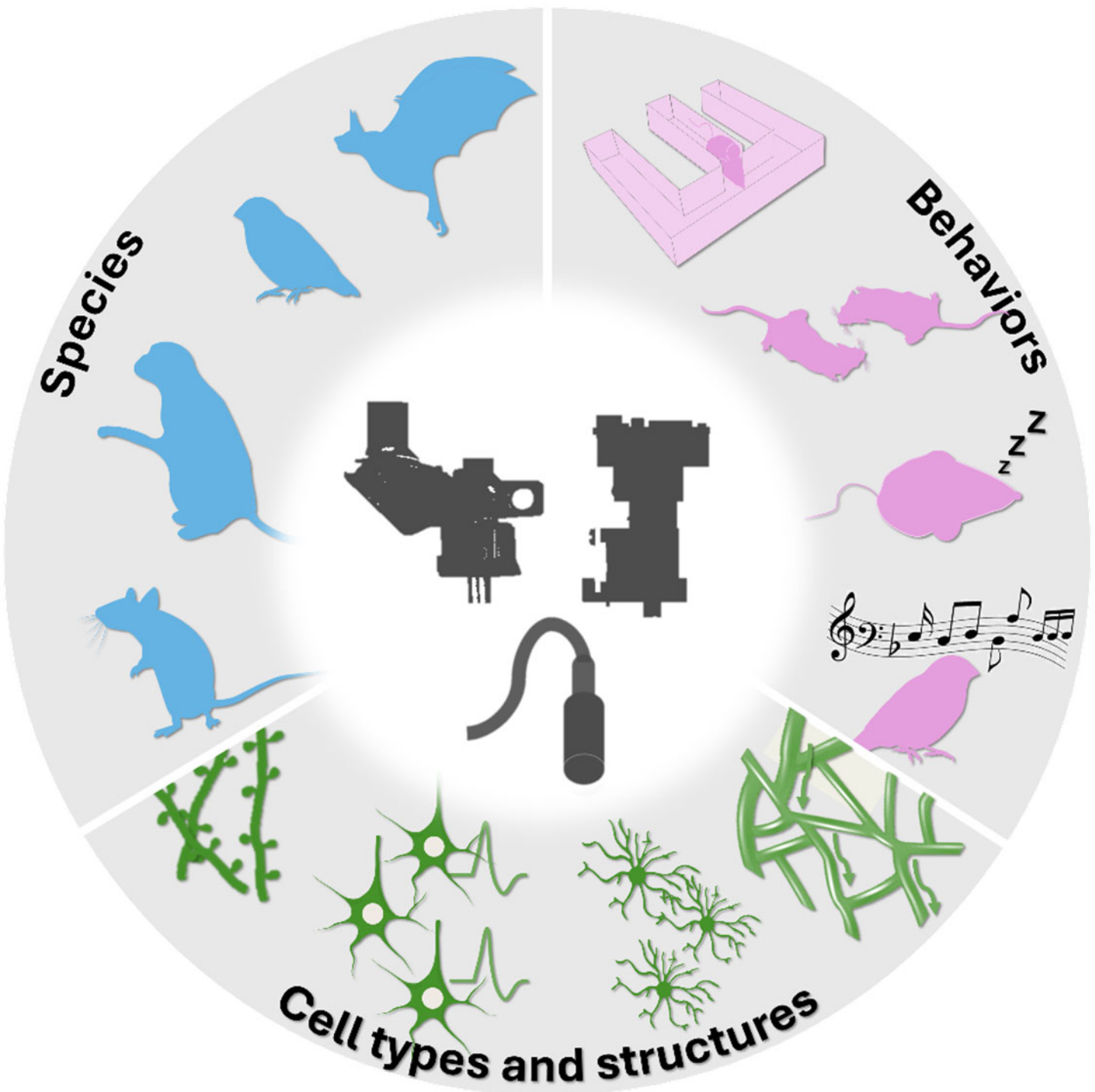

**Figure 1 | Applications of head-mounted miniscopes in neuroscience.**
Miniscopes enable recording of neural activity across diverse species (mice, rats, songbirds, bats, and non-human primates) during behaviors that are difficult or impossible to study in head-fixed conditions, such as spatial navigation, social interaction, sleep, and vocal communication. Combined with genetically encoded or vascular indicators, they allow imaging of neurons, astrocytes, microglia, dendrites, spines, and blood vessels. This versatility has made it possible to study neural dynamics at multiple scales, from subcellular compartments to local circuits and distributed brain networks in freely behaving animals.

## Basic architecture of miniscopes

Most miniscopes preserve the fundamental optical architecture of their benchtop counterparts but use miniaturized optical components (Fig. 2). This results in a compact form factor, typically weighing 3–4 grams and measuring less than 20 × 20 × 25 $mm^3$. In general, microscopes function by delivering excitation light to a sample and collecting the resulting scattered or emitted light, which carries information about the sample. The excitation and collection paths together form the backbone of the microscope's optical system. Under such a backbone framework, we classify miniscopes based on the volume of the sample that is illuminated by the excitation source (laser or LED) — either a widely-spread or confined volume, and whether a planar (i.e. pixel array such as a camera) or point detector is used to collect the signal (Fig. 2a).

Class A excites the sample over a large volume and detects the signal through a planar detector, represented by 1P epifluorescence miniscopes. The basic 1P miniscope illuminates the entire sample using blue or green LED light and images the resulting green or red fluorescence across the full field-of-view (FOV) simultaneously onto a camera sensor through an objective lens and a tube lens[9,15] (Fig. 2b-i). An electrically tunable lens can be incorporated into the collection path to adjust the focal depth. The wide-field excitation and parallel acquisition strategy simplify the optical design and allow for high-speed imaging over a large FOV. However, these systems are typically limited to imaging superficial layers in intact, scattering tissue with strong phototoxicity, and the collected signals are often contaminated by out-of-focus background fluorescence (Box 1). One approach to reject out-of-focus background fluorescence is through structure light illumination[11,57,58], where multiple periodic patterns are used in excitation, and computational algorithms are used to synthesize the image (Fig. 2b-ii). Here, the excitation light is patterned through benchtop optics, and then transmitted to the miniscope through a fiber packed with multiple cores (termed as imaging fiber bundle). The fluorescence can either be transmitted through the same fiber bundle back to the benchtop collection optics or detected directly at the camera within the miniscope.

Class B uses a completely different strategy, with confined excitation and point detector, represented by multiphoton miniscopes, which image a single voxel at a time (Fig. 2a). These systems scan the focus of an infrared femtosecond laser, delivered via a fiber, across the FOV, and excite the sample pixel by pixel. The resulting fluorescence is collected using a bucket detector. This approach, combined with the nonlinear nature of multiphoton excitation, enables deep tissue imaging with high lateral and axial resolution and a superior signal-to-background ratio, together with a reduced phototoxicity (Box 1). However, these advantages often come at the cost of a smaller FOV and substantially higher system complexity and expense.

Depending on the beam scanning mechanisms, multiphoton miniscopes can be implemented in various architectures. The first type, also known as a fiberscope, employs a double-clad fiber with a solid core and two layers of cladding to deliver the femtosecond laser, where the laser light is confined in the core[24,28] (Fig. 2b-iii). A piezoelectric device is attached near the tip of the fiber to form a fiber cantilever, which can be spirally scanned at resonant frequency. The emission at the fiber tip is relayed to the sample, and the fluorescence returns through the same optical path, collected by the core and inner cladding of the fiber. This method offers a small device footprint but has challenges for large FOV as the large bending angle of the cantilever could introduce substantial aberration. The second type mirrors the optical layout of its benchtop counterpart[29,33,34] (Fig. 2b-iv). It uses a hollow-core fiber to deliver the femtosecond laser to the miniscope. The fiber confines light within the hollow core to minimize nonlinear distortion to the femtosecond light pulses. A MEMS scanner, operating at resonant frequency on the fast axis, scans the beam across the sample. The focal depth of the excitation light is adjustable via an electrically tunable lens. The fluorescence is collected by another fiber and detected with a photomultiplier, or directly detected with an onboard silicon photomultiplier. The third type utilizes remote scanning at the bench (Fig. 2b-v). The imaging fiber bundle is used to deliver the laser light. Scanning is performed across the individual cores at the proximal end of the fiber through benchtop optics, and the resulting pattern is relayed to the distal end at the sample site[30]. The same fiber bundle can be used to collect fluorescence. This approach may further miniaturize the microscope but could suffer from cross-talk among different fiber cores, and the limited average power that can be used for imaging due to the nonlinear effect in the fiber core. One strategy to overcome this is to spread the laser over multiple cores (see class C below, Fig. 2b-vii). When solid core fibers such as the double-clad fiber or fiber bundle is used to deliver the femtosecond laser, sophisticated dispersion compensation is needed at the bench to compensate for the nonlinear distortion by the fiber. For 3P miniscopes, it requires excitation pulses with high peak power, imposing challenges on using solid core fiber for light delivery due to strong nonlinearity. So far, all the 3P miniscopes are implemented through MEMS scanner[37-39].

Class C is a hybrid technique, combining confined excitation with planar detection (Fig. 2a), represented by the miniscopes using light-sheet illumination (Fig. 2b-vi). Here, the excitation beam is shaped into a thin layer of sheet and a prism is used to direct the sheet orthogonal to the detection path[59]. Interestingly, one variation of the remote scanning approach in multiphoton miniscope also uses class C strategy, where the laser spot is spread to illuminate multiple cores simultaneously to decrease the power per core (Fig. 2b-vii). This increases the excitation area, and therefore a camera is used for detection to achieve high resolution[60]. One potential solution is to use a single bucket detector, which can mitigate crosstalk. However, implementing this strategy may require specific designs for both the excitation scheme and the image reconstruction algorithms (see Section Towards high-throughput miniscope below and Fig. 3). Class C techniques reduce the phototoxicity though is limited to shallow imaging depths in scattering tissues, as pixel crosstalk becomes substantial at greater depths.

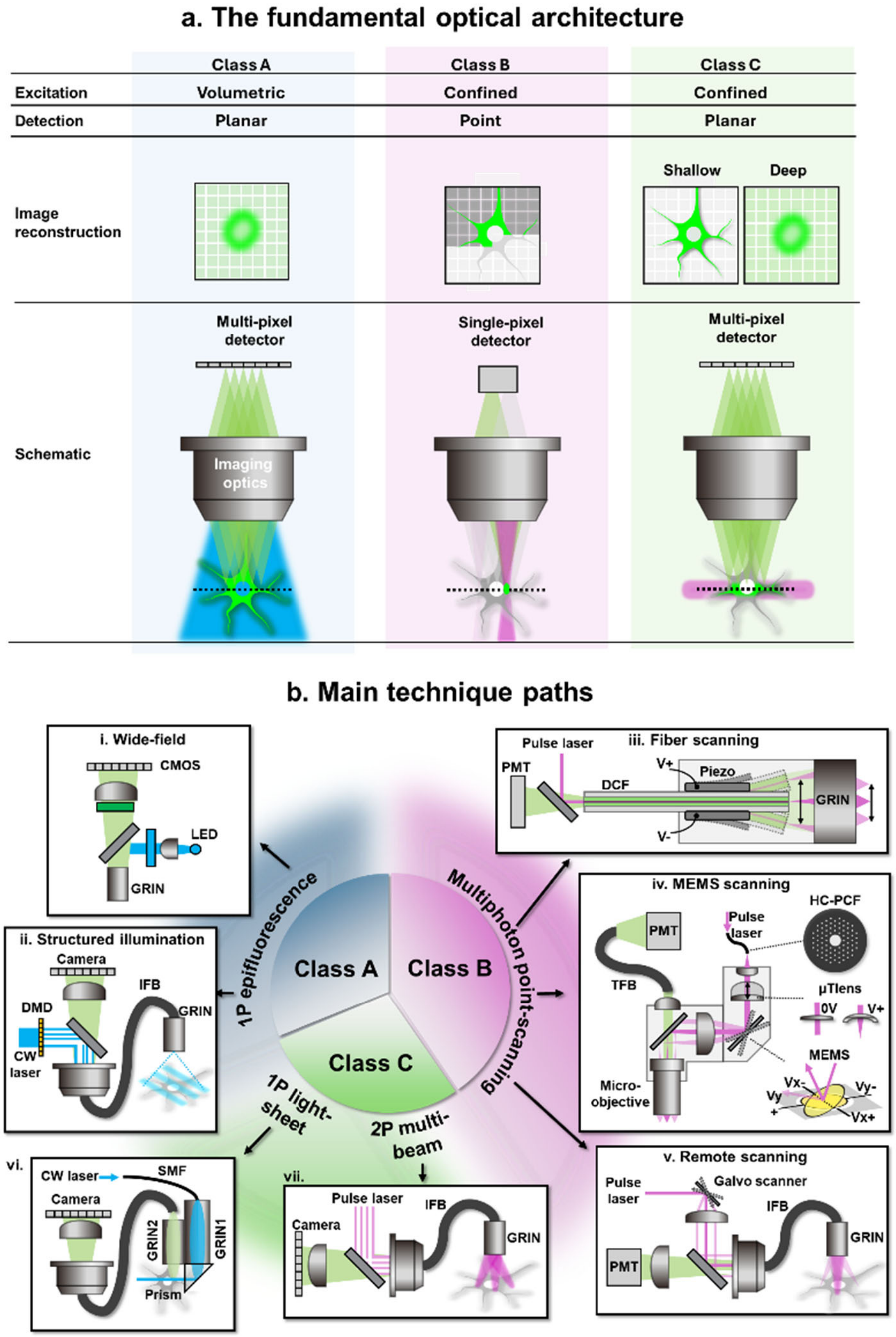

**Figure 2 | Core architecture and representative designs of miniscopes.**
**(a)** Conceptual classification of miniscope architectures. Class A uses wide-field illumination with planar detection (e.g., 1P epifluorescence miniscopes), offering fast imaging but with strong neuropil and out-of-focus contamination. Class B employs point excitation and single-pixel detection (e.g., 2P/3P point scanning miniscopes), providing high resolution and depth penetration at the cost of slower imaging speed. Class C combines confined excitation with planar detection (e.g., light-sheet or fiber-bundle–based systems), balancing speed and resolution in shallow tissue but with degraded image quality at depth.

**(b)** Representative optical layouts. Class A includes (**i**) basic 1P wide-field miniscopes and (**ii**) structured illumination designs using digital micromirror devices (DMDs) and imaging fiber bundles (IFBs). Class B comprises multiphoton devices with different scanning strategies: (**iii**) fiber scanning using a piezo-actuated, (**iv**) MEMS (miniaturized resonant mirrors) scanning, and (**v**) remote scanning (beam steered by bench-mounted scanners and relayed to the sample by IFBs). With MEMS-scanning strategy, the state-of-art systems applied hollow-core photonic crystal fibers (HC-PCF) to deliver femtosecond lasers and electrically tunable lenses (uTLens) to adjust imaging depth. Class C includes (**vi**) 1P light-sheet systems and (**vii**) multi-beam remote-scanning 2P variants that spread excitation across multiple fiber cores for camera-based detection.

---

**Box 1: Selection guide of 1P epifluorescence (class A) and 2P/3P (class B) miniscopes**

In 1P miniscopes, the entire FOV is illuminated and imaged at once (class A). Unless imaging the tissue surface, the fluorescence needs to traverse some thickness of the tissue before captured by the camera, during which it could get scattered, thus creating crosstalk between pixels on the image. The consequence is a blurred image, effectively reducing the resolution. Furthermore, the broad illumination excites the fluorescence over an extended volume in the sample, but the imaging path often has a limited depth of focus. The fluorescence generated outside the depth of focus becomes out-of-focus light, and appears as a broad background on the image. The deeper the imaging depth, the stronger the effect from the scattering and out-of-focus background. Furthermore, a larger cell labeling density will also increase the background.

To image deep brain structure with 1P miniscope, a common approach is to insert a graded index (GRIN) lens[9] or a compound element with a GRIN and a micro-prism[61] to the brain. The GRIN lens or GRIN-micro-prism compound element effectively extends the microscope and bypasses the superficial scattering tissue, though they could be highly invasive. Furthermore, by bypassing the layer 1 in the cortex, which is densely packed with dendrites and axons and thus not exciting or collecting their fluorescence, this method could significantly suppress neuropil signals from extracellular and out of focus regions.

As the minimum feature size that 1P miniscopes image is the cell body, somatic targeting labeling can be used to avoid labeling dendrites or axons, thereby reducing background[62-64]. Additionally, if the research question allows for sparse labeling, such as in studying interneurons, this can further reduce the background fluorescence.

Multiphoton excitation (class B) can image substantially deeper in the intact tissue, with a much reduced background and fine resolution[65-68]. As an example, while it requires the invasive GRIN lens to image the hippocampus with 1P miniscope, 3P miniscope can image it from the cortical surface with minimum invasiveness[39]. Multiphoton microscopes have the following features. First, it uses a long excitation wavelength (infrared) which is more resistant to light scattering and can deliver a tight focus deep into the tissue. Additionally, as the image is built pixel by pixel, the detector collects fluorescence from one pixel at a time. Though the fluorescence (visible wavelength) is scattered, it can still be registered to the corresponding pixel. Second, the nonlinear excitation scheme confines the fluorescence generation around the focal spot, thus substantially reducing the out-of-focus background. Third, multiphoton excitation also allows imaging different depth with fine axial resolution, thus allowing studying the interaction between different layers and also increasing the number of neurons that it can sample[33,34]. Finally, multiphoton miniscopes allow imaging fine features, such as dendrites and spines in highly scattering tissue[29].

Compared to 1P, multiphoton excitation has a smaller FOV limited by the maximum scanning angle that the scanner can achieve. Furthermore, the imaging speed is generally lower. In particular, 3P excitation requires high laser peak power due to its low excitation efficiency associated with the high-order nonlinearity. This necessitates the usage of a low repetition rate laser, which could further limit the imaging speed. Finally, 2P and 3P excitation requires expensive femtosecond lasers, though the laser cost could be reduced with the advancement of fiber laser technologies.

The table below shows the typical imaging parameters among 1P, 2P, and 3P excitation, which could be chosen based on the requirement of the research questions.

| | 1P | 2P | 3P |
|---|---|---|---|
| Imaging depth in scattering tissue from cover glass in cranial window, or end facet of GRIN lens or micro-prism | < ~300 μm | > 600 μm | > 1 mm |
| FOV | > 1 mm | <=1 mm | < 1 mm |
| Lateral resolution in scattering tissue | cellular | Sub-cellular | Sub-cellular |
| Axial resolution | NA | 5~20 μm | 5~20 μm |
| Imaging speed<br>(with pixel counts in 2P/3P) | 30-60 Hz | 10-40 Hz<br>(256x256) | 10-15 Hz<br>(300x300-128x128) |
| Background sensitivity to label density | High | Low | Ultra low |

---

**Towards the high-throughput miniscopes**

Similar to the benchtop microscopes, the miniscopes development aims to increase the imaging FOV, while maintaining a high spatial resolution and imaging speed, with an additional constraint of device footprint and weight (Fig. 3-4). These metrics could be condensed into a term called *imaging* throughput, defined as the product between *optical* throughput and imaging speed. The optical throughput, also interchangeably termed as etendue or optical invariant in literature[69], represents the number of resolvable points of the microscope, and is defined as the product between FOV and numerical aperture which is inverse proportional to lateral resolution (Box 2). In the past decade, different classes of miniscopes have rapidly increased the imaging throughput, with notable improvement of FOV in both 1P[14,17-20,22,70] and 2P devices[36,71] (Supplementary Table 1, 2). Here, we discuss the factors leading to these improvements and the challenges that lie ahead.

---

**Box 2. Optical invariant.**

Optical invariant, calculated with the height and angle of the chief and marginal rays, is a measure of light propagating through an optical system[69]. In a system without aberration and vignetting, optical invariant is conserved as the rays propagate along each optical component. The optical invariant can be defined separately for the excitation and collection path. In a 2P miniscope, considering the beam radius at the MEMS scanner and objective lens back aperture as $r_m$ and $r_p$ respectively, the maximum scanning angle at these two planes as $\theta_m$ and $\theta_p$ respectively, and the radius of FOV and excitation numerical aperture (NA) on the sample plane as $r_f$ and $NA_{ex}$ (which can be expressed as the product between the refractive index $n$ of the medium between the objective lens and the sample, and $sin\theta_e$ with $\theta_e$ being the half-cone angle of the excitation light after the objective lens) respectively, the optical invariant of the excitation beam $I_e$ can be expressed as[72]

$$I_e = r_m sin\theta_m = r_p sin\theta_p = r_f NA_{ex} = r_f n sin\theta_e$$

The significance of $I_e$ becomes clear as $r_f NA_{ex}$ is proportional to the ratio between FOV and lateral resolution, or the number of resolvable points that can generate fluorescence. For the same $I_e$, there is a tradeoff between FOV and resolution. Increasing both FOV and resolution necessitates an increase of $I_e$.

At the collection path, the optical invariant of the collection beam[72]

$$I_c = r_f NA_c = r_d sin\theta_d = r_d NA_d$$

where $NA_c$ is the collection NA of the objective lens, $r_d$ and $\theta_d$ are the beam radius and maximum beam incident angle at the detector respectively, and $NA_d$ is the effective numerical aperture of the detector.

In class B multiphoton miniscope using a point-wise excitation and collection strategy (e.g. multiphoton miniscope), $I_e$ directly influences the spatial resolution and the FOV over which a fluorescence signal can be generated, and $I_c$ is proportional to how much of the generated fluorescence can be captured. While increasing both $I_e$ and $I_c$ is desirable, $I_e$ is generally the more challenging factor to optimize.

In contrast, class A 1P miniscopes use wide-field illumination, so only $I_c$ is of interest. Here, $I_c$, which is measured at the camera detector, governs both the spatial resolution and the FOV.

For 2P miniscopes that use a camera for image acquisition (class C), both $I_e$ and $I_c$ influence the FOV, while the spatial resolution is determined by $I_c$.

In practice, $I_e$ and $I_c$ is determined by the component that supports the smallest optical invariant in the entire system. Increasing $I_e$ and $I_c$ requires a holistic consideration of all components, and typically involves an increase of the aperture (i.e. NA of the objective lens, or the size of the camera) or the angle of the marginal rays (e.g. the scanning angle of the scanner). This often results in a larger device footprint or a reduced imaging speed.

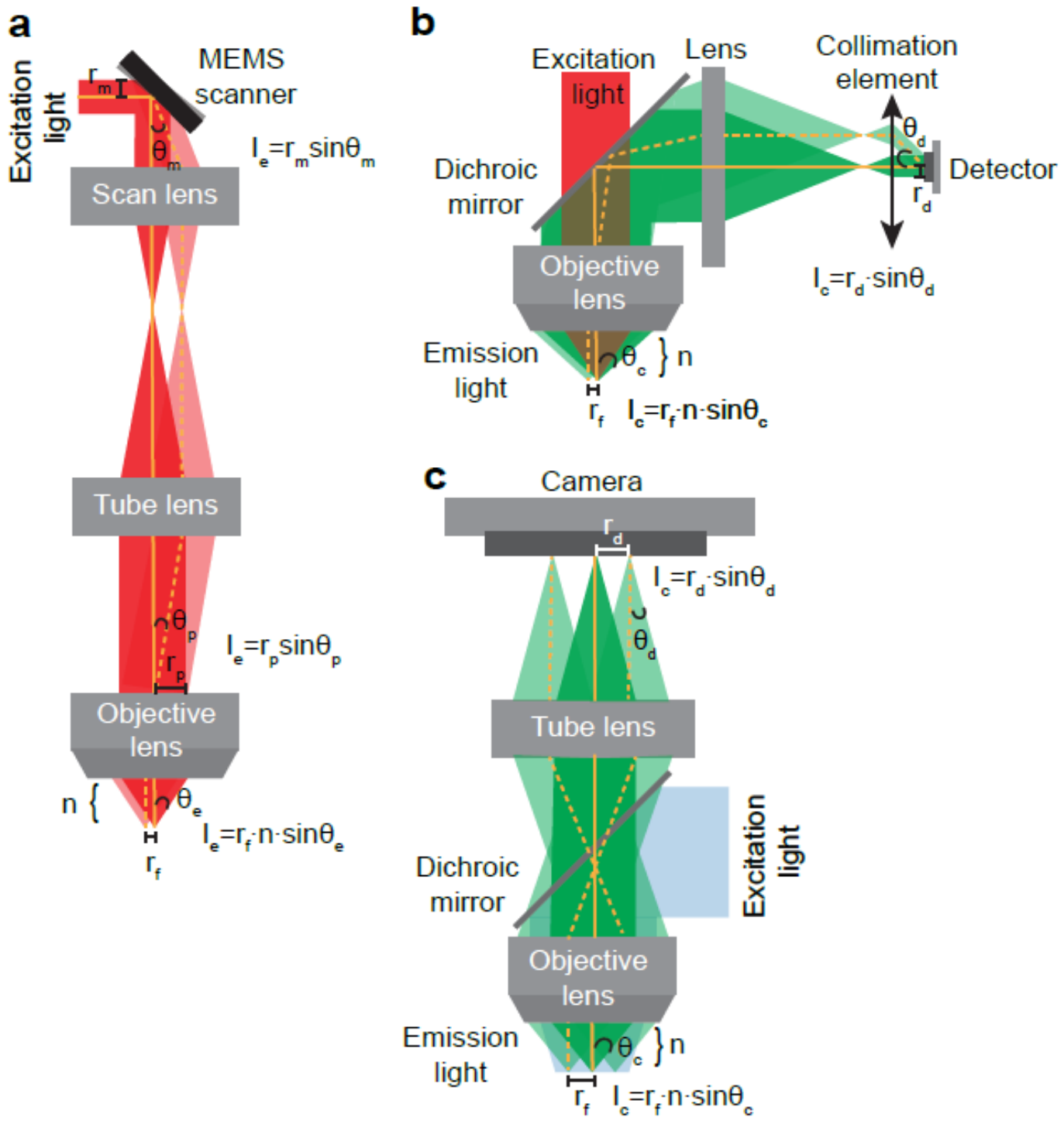

**Box 2 figure |** Illustration of the optical invariant for (a) excitation path of 2P microscope, (b) collection path of 2P microscope, and (c) collection path of 1P microscope. In 2P miniscopes, the tube lens and objective lens could be combined as a single objective lens so the Fourier plane in between them may not be accessed.

***Optical throughput improvement by engineering basic optical components***

In class A miniscopes, such as the 1P epifluorescence, sample illumination is typically not a limiting factor. Therefore, the optical invariant of interest lies in the collection path ($I_c$) and is governed by the lens design and the camera. Ultimately, the number of pixels in the camera sensor sets the upper limit for the number of resolvable points and thus $I_c$. Over the past years, advancements in the semiconductor industry have substantially increased the pixel density and resolution of camera sensors used in 1P miniscopes (Supplementary Table 1). The pixel count of these sensors has risen from ~0.3 million in early generations[9,12] to ~5 million in the latest generations[18,20,22,70]. Concurrently, the pixel size has decreased from ~6 μm[9,12] to 1~2 μm[18,20,22,70], and the diagonal size of the sensor has grown from ~5 mm[9,12] to ~7 mm[18,22,70]. This trend is coupled with a reduction in optical magnification from ~5-6x[9,12] to ~1x[18-20,22,70] between the brain sample and the camera sensor. Consequently, this results in a large FOV, while the smaller pixel size ensures that images of individual neurons are covered by a sufficient number of pixels to maintain proper image quality and signal-to-noise ratio (SNR). While new generations of large-FOV miniscopes could continue to use the basic optical architecture with an objective lens, tube lens, and a dichroic mirror in between to separate excitation and emission paths (Fig. 3a), and be built using off-the-shelf components[22], new optical architectures have been developed to reduce the device's footprint and weight, while supporting a large $I_c$. In one example[19] (Fig. 3b), two sets of precision molded plastic aspheric lens assemblies, originally developed for smartphone cameras, provide a 1:1 relay between the sample and camera sensor. Excitation light is delivered through a fiber ending with a 90$^o$ prism, placed between the lens assemblies. In other examples, a single custom-designed diffraction optical element[70] (Fig. 3c) or a set of custom designed aspherical lenses with a phase plates[20] serve as the imaging optics. These optics have a small lateral size, but can support an FOV larger than themselves. Furthermore, they can image over an extended depth of field. The sample is illuminated at an angle by LEDs mounted at the peripheral of the optics.

At present, the 1P miniscope can reach an FOV larger than 3.5 mm with a lateral resolution of 3-4 μm[18-20,22,70] (Fig. 3a-3c). FOV of 10 mm is also demonstrated though the resolution is degraded to >10s μm[17] (Supplementary Table 1, Fig. 3d). While camera sensors with large pixel count generally have a reduced frame rate, new generations of camera sensors with 5 million pixels can already reach 60 Hz imaging rate, enough for calcium imaging. It is envisioned that $I_c$ of the 1P miniscopes can be further increased with the development of high-speed, large pixel-count camera sensors, coupled with optics designed to have a <1x magnification and support a large optical invariant. As the pixel size becomes smaller, the SNR per pixel could drop, though this may be recovered through image processing technique. Another challenge is that the printed circuit board hosting the camera sensor still has a relatively large footprint, or is not commercially available in smaller footprints due to low demand. However, it is anticipated that these boards will become smaller with increased adoption of large-FOV miniscopes. An interesting direction is to employ multiple lightweight miniscopes to simultaneously image spatially separated FOVs in the brain[16,21,43,73]. For example, two NINscopes (1.6 g each, in comparison with ~2.0 g in commercialized system) were used to image two distinct regions[16], while more recently, four ultra-light TINIscopes (0.43 g each; dimensions: 4.6×13.8×8.24 mm$^3$) were deployed to monitor four separate regions[21] (Fig. 3e). Additionally, the mini-MCAM, a miniaturized micro-camera array microscope utilizing four micro-endoscope cameras (1.175 g each) further extends these capabilities by imaging the dorsal cortex with an expansive FOV of 4.5x2.55mm$^2$ per camera, though its resolution is degraded to be 10-20 μm and its extensive cabling currently restricts the animal's motion[73] (Fig. 3f).

Unlike the class A 1P miniscope where the development is centered around the collection path, in the class B point-scanning 2P miniscope, the major goal is to deliver the excitation spot with fine resolution to the sample over a large FOV (thus a high $I_e$), and scan it as fast as possible. In the early generations of 2P miniscopes, the optical invariant of the excitation path ($I_e$) was limited by the objective lens, which often used off-the-shelf graded index (GRIN) lens that exhibited strong aberrations as the FOV increased. In the latest generations, the GRIN lens has either been custom-designed or replaced with custom-designed objective lens[33,34,36,71], which substantially increases the FOV while maintaining a high spatial resolution (Fig. 3h). Now the scanner becomes the limiting factor for $I_e$. In miniscope with a MEMS scanner, $I_e$ is determined by the product of its mirror size and maximum scanning angle. Through a 4f system formed by the scan lens and the tube lens with a magnification of *M*, the beam size expands by a factor of *M* while the scanning angle shrinks by a factor of *M* at the back aperture of the objective lens. Assuming the objective lens is not a limiting factor, a larger *M* results in a higher excitation NA and thus spatial resolution but a smaller FOV, and vice versa. Increasing the FOV while maintaining a high spatial resolution requires a larger $I_e$, i.e. a larger MEMS mirror size and/or higher scanning angle. Unfortunately, this usually reduces resonant frequency and thus scanning speed of the scanner, making it challenging to increase the overall imaging throughput. One approach to overcome this limitation is FOV stitching, achieved by engineering a spatial mounting mechanism that allows repositioning of the miniscope relative to the imaging window[34] (Fig. 3i). Maximum 2.5x2.5 mm$^2$ stitched FOV in the visual cortex has been shown using this method[34] (Fig.3i lower part). However, this method requires sequential imaging of different FOVs, which prevents simultaneous measurement and limits the study of temporal correlations between neurons across these regions.

In 2P miniscopes with a piezo-actuated fiber cantilever, the same tradeoff exists[31]. The fiber tip is conjugate to the sample plane through a 4f system with a magnification of $M^{-1}$. Assuming optics are not limiting, $I_e$ is conserved from the fiber tip to the sample plane, where both spot size and scanning range are shrunk by *M*. Increasing the cantilever length expands the scanning range but reduces the resonant frequency and thus the imaging speed. A longer, thicker piezoelectric tube increases the cantilever's angular deflection range and FOV but may not suit miniaturization. Moreover, larger deflections can shift the fiber tip away from the conjugate imaging plane and distort the symmetricity of the illumination cone with respect to the optical axis, which could introduce strong aberrations. One elegant way to increase $I_e$ while maintaining the scanning speed is to engineer the fiber tip by attaching a beam expander and focusing optics to it, which increases the NA of the light emitting from the cantilever and thus reduces the spot size[31] (Fig. 3g). This approach generally requires engineering to account for chromatic effects between excitation and fluorescence emission, as the same fiber cantilever collects the fluorescence emission.

The 3P miniscopes share the same optical layout and thus similar tradeoff in spatial resolution, FOV and imaging speed in the excitation path, as their 2P counterpart (with MEMS scanner). However, because 3P imaging targets greater tissue depths, the resulting increase in scattering of fluorescent emission makes efficient signal collection equally critical. This highlights the need for high NA optics in the collection path to maximize collection efficiency and ensure a high $I_c$.

The board-level silicon photomultiplier (SiPM)[38] (see the Section Emerging new technologies, Fig. 4b), which eliminates the collection fiber, can enhance the collection efficiency (Fig. 3j). One additional challenge in 3P miniscope is the low imaging rate, which stems from the low repetition rate laser required to achieve high peak pulse energy and compensate for the low 3P absorption cross-section while maintaining a low average laser power. Brighter and more sensitive fluorescent indicators, as well as optimized collection optics[39] (Fig. 3k) could help mitigate this constraint.

In Class C 2P miniscopes that utilize a fiber bundle to relay spatially extended excitation spots to the sample and transmit the resulting fluorescence to the camera, the fiber bundle often becomes the limiting factor in optical throughput. These bundles typically offer high numerical aperture (NA), and recent advancements have increased the overall bundle diameter, thereby expanding the FOV and enhancing the fluorescence collection efficiency[74] (Fig. 3p). To maintain flexibility, the outer cladding of each fiber core is removed. Nonetheless, to avoid interfering with the animal's natural behavior, the total diameter of the bundle must be constrained, which in turn limits the achievable FOV. Additionally, minimizing inter-core crosstalk is essential for enhancing spatial resolution.

Advances have also been made in Class C light-sheet miniscopes, where thin photonic probes equipped with optical waveguides and grating couplers are used to deliver the light sheet. Individual light sheets emitted from multiple grating couplers can be seamlessly merged to form a continuous light sheet, thereby expanding the FOV. Additionally, light sheets can be directed to different depths, and depth-resolved imaging can be achieved through a multicore fiber combined with computational reconstruction algorithms[75,76] (Fig. 3o).

In general, microscopy with a larger optical invariant requires a larger aperture and more complex design of optics to account for the increased aberration. The optical lenses generally have a larger size in 3D, where miniaturized microscopy cannot afford. Using aspherical lenses rather than the conventional spherical lenses, as already seen in 1P miniscopes[19,20], could substantially reduce the size in the axial direction. This may benefit 2P/3P miniscopes as well though the cost is substantially higher.

### *Imaging throughput improvement in multiphoton miniscopes by beam multiplexing or PSF-engineering*

The point-wise imaging nature of multiphoton microscopes inherently limits their imaging speed. In the past two decades, various strategies, such as beam multiplexing and PSF engineering, have been developed in 2P benchtop microscopes to increase the imaging speed and thus imaging throughput[77-85]. These techniques essentially excite the sample in an extended volume at a time, and capture the signal through a bucket detector. Similar principles could be used in the miniaturized version. The beam multiplexing approach utilizes multiple scanning beams[77-81] (Fig. 3l). Each beam scans a different spatial sub-FOV, with their pulse trains temporally delayed relative to each other. The signal at the detector is temporarily gated and assigned to the corresponding beam or sub-FOV. This method maintains the same overall FOV and spatial resolution, while increasing imaging speed by $N$ folds, where $N$ is the number of beams. Alternatively, different beams could image different depths. In another variant termed computational multiplexing[84], the pulse trains of individual beams can be aligned in time, causing recordings of individual sub-FOVs to overlap spatially. The activity of individual neurons in each sub-FOV can be demixed and extracted[84,86], given prior knowledge of their spatial footprint in the corresponding sub-FOV, which can be obtained by activating only one beam at a time as a calibration step[84]. These multiplexing schemes have been recently demonstrated in both 2P and 3P miniscopes, where *N* excitation beams were delivered to the same miniscope via *N* fibers (N=2~5)[55,87]. In one of these demonstrations[55], voltage imaging was realized in a freely behaving mouse at 400 Hz over an FOV of 380x150 μm$^2$. Given the thinness of each fiber, it is likely that even more fibers may be used to further increase the imaging throughput.

PSF engineering shapes the PSF into specialized forms. For instance, a Bessel beam PSF can be generated by creating an axicon lens at the tip of the delivery fiber, allowing an extended depth of field to be illuminated simultaneously[88] (Fig. 3m). This method shares similarities with the computational multiplexing described above, but the beam shape now extends continuously over a range in the axial direction. Consequently, a 3D volume can be imaged in a single frame scan. Given the spatiotemporal sparsity of the neuronal activity, their activity can be demixed, and the depth information can be obtained by imaging the same volume with a conventional benchtop 2P microscope. In another example, the PSF is shaped as a short line (5~10 μm) along the slow scanning axis in the in-plane direction through two orthogonal cylindrical lenses[89] (Fig. 3n). This method reduces the number of scanning lines and thereby increases the imaging speed. Essentially, each single measurement samples a larger tissue area or volume. As the length of the line is less than the cell diameter, cellular resolution can be maintained.

In general, increasing imaging speed is associated with higher laser power on the sample, raising concerns about thermal damage. The use of low-repetition rate lasers and brighter fluorescent indicators can reduce the laser power required to achieve the same SNR. Additionally, the application of imaging denoising algorithms allows for information recovery from low SNR recordings obtained with lower laser power[90,91].

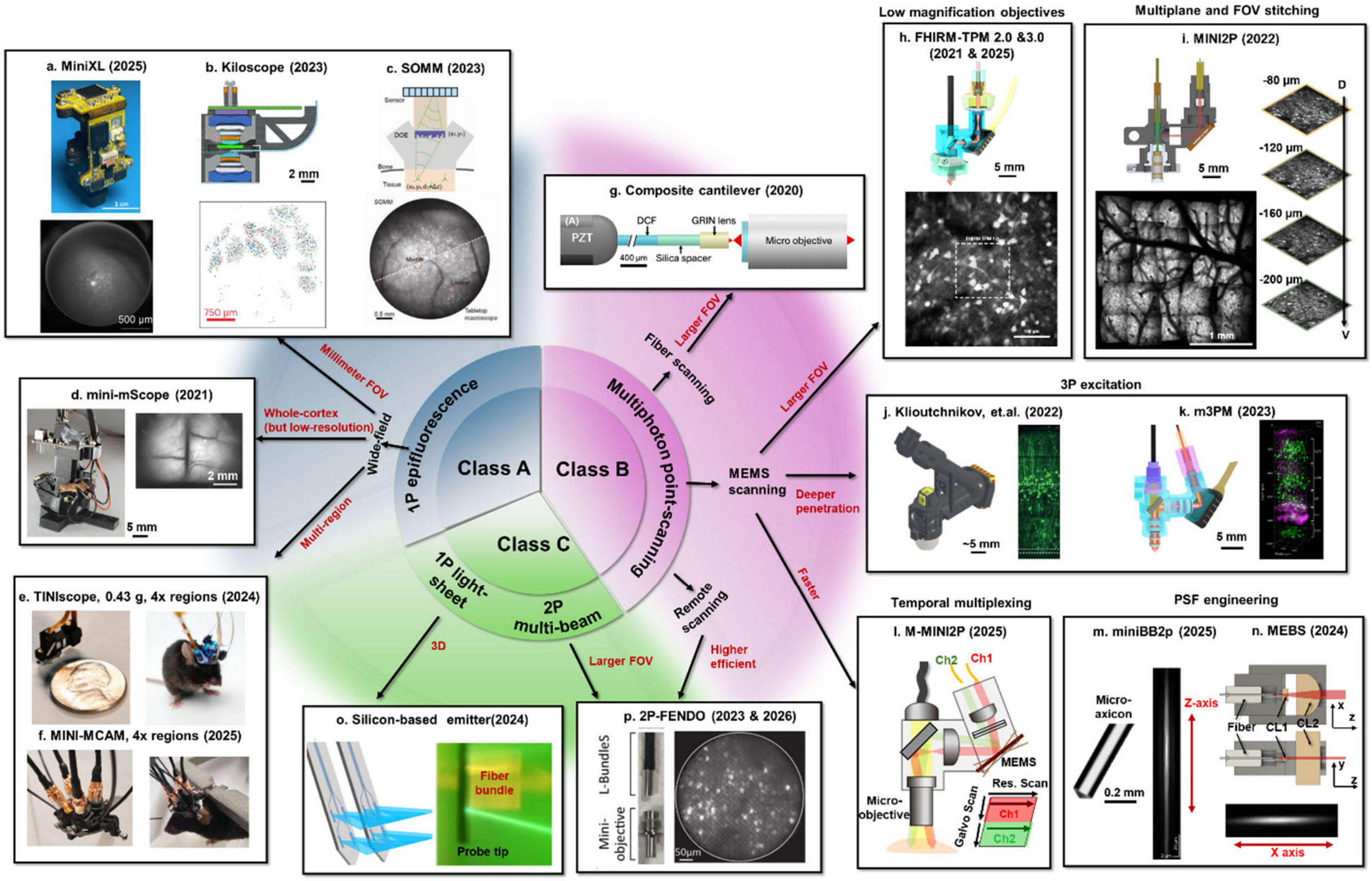

**Figure 3 | Featured development by engineering better optical components in the last 5 years.**
(**a-c**) In Class A, recent devices extend the field of view (FOV) to the mesoscale (a few millimeters), while maintaining single-cell resolution in sparsely labeled samples. (**d**)The brain-wide (>10mm) FOV with lower resolution (>10s μm) has also been achieved. (**e** and **f**) Moreover, ultralight miniscope designs have enabled (**e**) 4-region (TINIscope) and (**f**) 4-region mini-MCAM simultaneously recording. (**g**) For increasing the FOV within fiber scanning category in class B, a composite cantilever was developed to engineer the fiber tip by attaching a beam expander and focusing optics to it. (**h**-**n**) In class B, the development of the MEMS-based 2P systems appears in three major directions, a continue increasing of FOV using (**h**) low magnification objectives and (**i**) applying multiplane imaging and FOV stitching, an increase of penetration depth by three-photon excitation (**j** and **k**), and an increase in imaging speed by (**l**) beam multiplexing and (**m** and **n**) PSF-engineering. In Class C, new developments include (**o**) thin photonic probes with silicon grating-based light-sheet delivery and light field detection to achieve 3D imaging. There is also a new series of fiber-bundle–based multiphoton systems (**p**) with increased FOV and efficiency, classified between Class B and C because it uses fiber-bundle-camera-based planar detection relays with multi-beam multiphoton excitation. Figures were adapted from original publication with permission: a from Ref. [22], b from Ref. [19], c from Ref. [70], d from Ref. [17], e from Ref. [21], f from Ref. [73], g from Ref. [31], h from Ref. [33], i from Ref. [34], j from Ref. [38], k from Ref. [39], l from Ref. [55], m from Ref. [88], n from Ref. [89], o from Ref. [76], and p from Ref. [74].

## Emerging new technologies

Beyond the conventional optical layout and hardware, recent years have seen the emerging of new architectures and components for miniscope, which either substantially enhance or show strong promise in enhancing imaging performance (Fig. 4). There is also growing interest in integrating imaging with optogenetics, as well as multimodal imaging. We discuss these advancements in this session.

### *Computational 1P miniscopes for 3D imaging over large volume*

In conventional 1P miniscopes, the point spread function (PSF), which describes how a single point source is mapped to the imaging plane, is an Airy disk that is spatially localized. This allows direct visualization of neuronal footprints from raw recordings. In contrast, computational 1P miniscopes (Fig. 4a) employ more complex PSF patterns and rely on deconvolution algorithms to reconstruct the original object. They can support advanced capabilities like single-shot 3D imaging. By compressing high-dimensional information (e.g. 3D) into a 2D image and leveraging the sparsity of the object, these methods computationally recover the original 3D structure, effectively increasing imaging throughput. One example is the light-field miniscope[92] (Fig. 4a-i), which uses a microlens array at the imaging plane and positions the camera at its focal plane. This device records both spatial and angular information, enabling depth extraction. Imaging an axial range of ~360 μm in the mouse cortex at once and distinguishing neurons separated by ~15 μm was reported. Alternatively, the microlens array could also be placed at the Fourier plane of the objective lens[93] (Fig. 4a-ii). The focal lengths of different microlens units could vary to enhance the uniformity of the resolution across depth[93]. Another example is mask-based miniscopes[94-98], which uses a single thin mask (microlens array or other patterns) between the object and camera (Fig. 4a-iii). The PSF spreads spatially, and sources at different lateral locations can be imaged by different parts of the mask. The FOV is thus scalable as the lateral size of the mask and camera sensor chip increases, while the device thickness is maintained. Furthermore, at different object depths, the PSF scales laterally while staying in focus. This enables single-shot 3D imaging where computational algorithms can be used to recover object information. A major challenge is the generally high image background due to the extensively spreading PSF, particularly when the object is dense, though advanced computational algorithms may mitigate this. It is reported that multiple axial depths can be resolved over ~300 μm in mouse cortex, with an FOV > 1 $mm^2$ [98].

### *1P miniscopes based on multimode fiber for deep brain imaging*

One-photon deep brain imaging typically uses GRIN lenses, which have large diameters and can cause significant tissue damage. Multimode fibers offer a promising alternative: they relay light between excitation/collection optics and tissue like GRIN lenses, but with a much smaller diameter for a given imaging field of view[99,100], thus reducing tissue damage. Compared to fiber bundles, they also provide higher spatial resolution. By modulating excitation light at the proximal end, a focal spot can be holographically generated and scanned at the distal end, enabling point-scanning imaging. This requires precise calibration of the light field mapping between the two ends. A 110 µm multimode fiber has successfully imaged deep brain regions in anesthetized mice[100]. However, applying this technique to freely moving animals, with a long multimode fiber connecting the benchtop beam forming optics and the mouse brain, remains challenging, as the mapping is highly sensitive to fiber bending. Overcoming this will require active mapping correction[101] or engineering the fiber's refractive index profile to make it robust for bending[102].

***Hardware improvement in multiphoton miniscopes***

In addition to the optics, other components in the multiphoton miniscope system, including the detector, delivery fiber, optical commutators and laser have evolved substantially in recent years. To detect fluorescence, the typical approach has been to collect it through a fiber and detect it with a photomultiplier tube (PMT) at the bench. Though compact PMTs could be mounted directly onto the miniscope body, they remain relatively bulky[7] or sub-optimal in efficiency[7,103]. Recently, a miniaturized, board-level silicon photomultiplier (SiPM), which is a single-photon avalanche diode array, has been fully integrated into the miniscope body, offering a more space-efficient solution[38] (Fig. 4b). This integration not only increases detection efficiency due to a reduced signal loss (and thus effectively increases the optical throughput in the collection path) but also replaces the collection fiber with a much thinner electrical wire, enhancing the animal's ability to move freely. Furthermore, compared to PMTs, SiPMs are more robust against damage by ambient light exposure, though it remains critically important to prevent ambient light or excitation light leaking into the SiPM to maintain signal integrity. SiPMs have a much higher dark current than PMTs. Despite this, they can be advantageous for fast imaging applications where the brief exposure time per pixel limits the accumulation of dark current counts.

Hollow-core fiber (HCF) is ideal for delivering femtosecond lasers, as it minimizes light interaction with material and thus reduces pulse distortion. Photonic bandgap HCF, which uses a photonic crystal structure to confine light in the air core, was first adopted in miniscopes[26]. Recently, anti-resonant type HCF has been optimized for broadband operation across 700-1060 nm, offering lower transmission loss and dispersion while maintaining a similar diameter to the photonic bandgap type[36] (Fig. 4c). Multiple excitation wavelengths can be transmitted through the same HCF to image structures labeled with different fluorophores[36]. However, compared with photonic bandgap HCF, anti-resonant type HCF generally has a smaller NA, and a higher bending loss.

Commutators are important elements in miniscope for twist-free operation to enhance the device robustness and applicability for natural behavior. While electrical connection over rotation is readily resolved through slip-rings, it is non-trivial to engineer a robust optical commutator for fibers delivering femtosecond lasers, as the nonlinear excitation is particularly sensitive to power fluctuation. Recently, such an optical commutator is demonstrated in the 2P fiberscope[32] (Fig. 4e). The commutator is composed of a pair of collinearly aligned fiber collimators, with one mounted stationary and the other connected to a rotary shaft. An encoder measures the torque of the fiber and a motor drives the shaft to compensate. With such a commutator, laser power fluctuation is kept within ±1%[32]. It is expected that such a technology will be adopted in multiphoton miniscopes with MEMS scanners, particularly when the collection fiber is no longer needed and replaced by SiPM.

Finally, femtosecond laser sources are becoming cheaper and more compact. Conventional 2P miniscopes use tunable Ti:Sapphire lasers. Recently, high-power femtosecond fiber lasers at 920 nm have been developed. Although not tunable, these fiber lasers are lower in cost and have a much more compact footprint than Ti:Sapphire lasers, allowing the entire system to be placed on a movable cart. Additionally, their repetition rate can be engineered to be lower than the standard 80 MHz, which could enhance 2P excitation efficiency and effectively reduce the average light power on the tissue. However, such fiber lasers still account for more than half of the total cost of 2P miniscopes. An even lower-cost laser system is desirable.

***Integration with precision optogenetics***

Optogenetics[104,105] is a powerful tool to investigate causal relationships among neural activity, circuit dynamics, and behavior, and it holds potential for treating brain disorders. Integrating imaging and optogenetics in miniscopes allows simultaneously imaging and manipulation of the neuronal activity, while observing the animal's behavior in real-time as it freely moves. To avoid crosstalk between imaging and photostimulation, calcium indicators and activity actuators (i.e., opsins) must have different excitation spectra, requiring two light sources with different wavelengths. Optogenetics has been integrated with 1P miniscopes by adding a second channel of excitation light. However, the general approach lacks spatial specificity, as the excitation light broadly illuminates the sample. Recently, digital micromirrors have been integrated into 1P miniscopes to pattern the 1P photostimulation light, allowing user-selected neurons to be stimulated[106] (Fig. 4d-i). Near-cellular resolution for the photostimulation beam was validated with 10 µm lateral resolution and 30 µm axial resolution in 200 µm brain slices. Alternatively, 1P pattern optogenetics can be achieved by a benchtop spatial light modulator and a fiber bundle to deliver the patterned light to the miniscope[11]. As a 1P technique, cautions should be taken as the out-of-focus background may bias the neuronal circuit, particularly when the number of simultaneously stimulated neurons increases.

Compared to 1P, 2P optogenetics can substantially increase both spatial specificity and accessible depth. While the optogenetics path can be implemented through MEMS, the imaging core fiber bundle approach is advantageous as it offloads beam scanning (via galvanometer mirrors) or patterning (via a spatial light modulator) to the bench, saving space in the head-mounted microscope. In one demonstration, the same fiber bundle was used for delivering excitation light for both the imaging laser and photostimulation laser, as well as for collecting fluorescence light[60,74] (Fig. 4d-ii). In another demonstration, the imaging path was realized with a MEMS scanner and a separate fiber for delivering the excitation laser[107] (Fig. 4d-iii). In both cases, the photostimulation laser was patterned into disks to match the shape of neurons, and allowed multiple neurons to be stimulated simultaneously. An interesting phenomenon is that the inhomogeneous structure of individual cores in the fiber bundle creates a temporal delay among different cores. In other words, within a disk pattern, pulses at different locations arrive at slightly different times (within a few ps). This effectively achieves temporal focusing, which tightens the axial confinement of the excitation light even though it extends laterally. To maintain high excitation efficiency while using low average power, a low repetition rate laser (1–10 MHz) was used in the photostimulation path. The high peak power creates strong nonlinearity in the fiber bundle and distorts the pulses, necessitating sophisticated temporal pulse shaping techniques to compensate for the pulse distortion.

### *Multimodal imaging*

While the miniscope was initially developed to record neural activity, it has the potential to support various imaging modalities. For example, with multiple excitation wavelengths, multimodal 1P miniscopes were demonstrated to record neural activity, hemoglobin absorption through the intrinsic optical signal and cerebral blood flow through the laser speckle contrast[108] (Fig. 4f-i). A recent study showcased the integration of laser scanning confocal fluorescence and photoacoustic microscopy, where a miniaturized ultrasound transducer is placed below the objective lens, enabling simultaneously imaging of neural activity and hemodynamics in freely-moving mice[109] (Fig. 4f-ii).

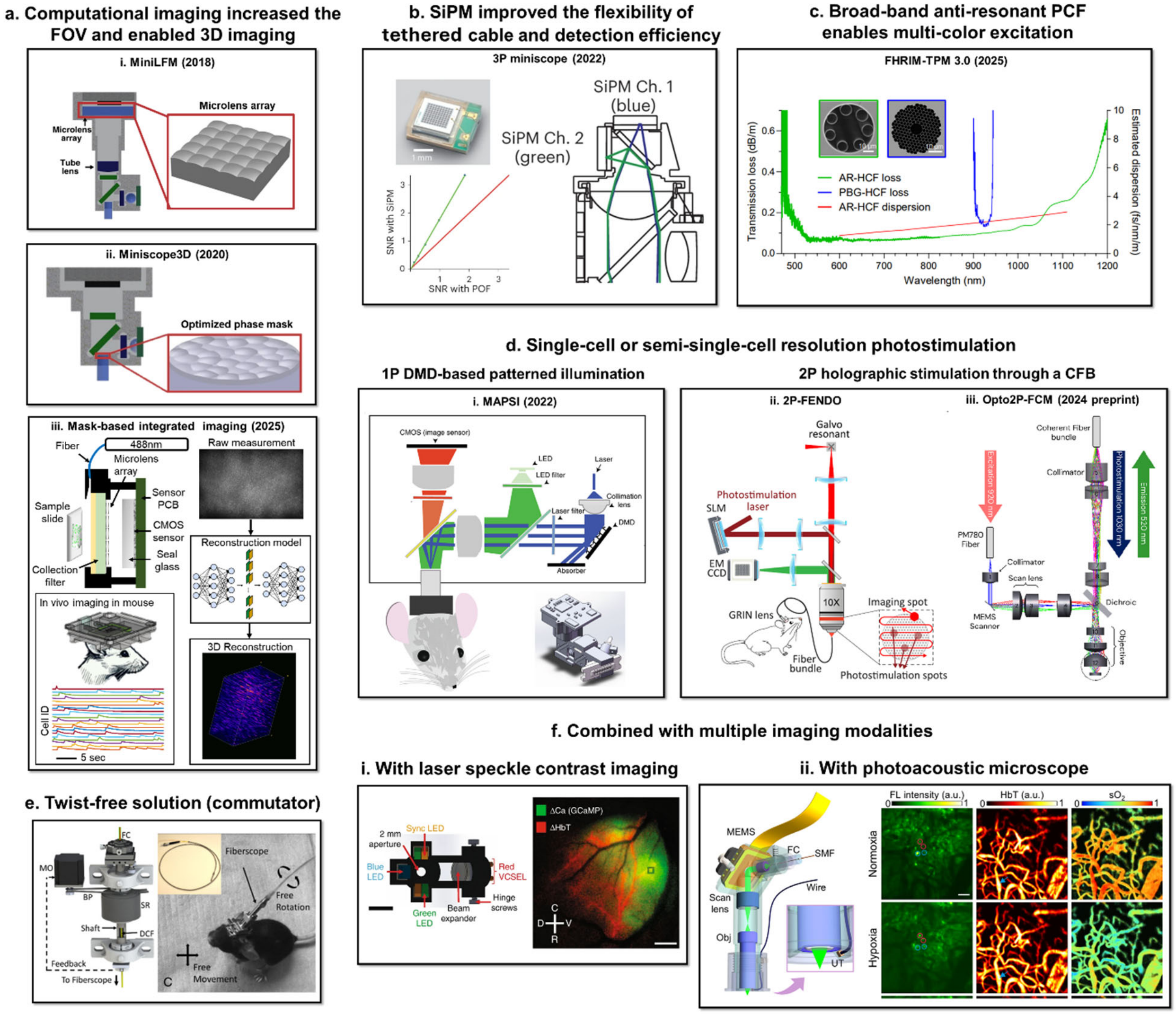

**Figure 4 | Emerging technologies expanding the capabilities of head-mounted miniscopes.**

**(a)** Computational imaging approaches that rely on engineered point spread functions and computational reconstruction.

**(b)** Integration of on-board detectors such as silicon photomultipliers (SiPMs) improves fluorescence detection efficiency, eliminates bulky collection fibers, and enables multichannel recording.

**(c)** Advances in hollow-core fibers (HCF), including anti-resonant HCF (AR-HCF), increased transmission bandwidth, allowing multicolor multiphoton excitation. PBG-HCF, photonic bandgap photonic crystal fiber (same concept as HC-PCF).

**(d)** Integration of (**i**) 1P and (**ii** and **iii**) 2P patterned optogenetics in miniscopes enables simultaneous recording and targeted manipulation of neuronal ensembles in freely behaving animals. SLM, spatial light modulator.

**(e)** Engineering of optical-electronic hybrid commutators supports twist-free animal movement with minimal loss of optical signals, facilitating long-duration behavioral studies. FC, fiber collimator; MO, motor; BP, belt and pulley; SR, slip ring, and DCF, double-clad fiber.

**(f)** Multimodal miniscopes expand beyond calcium imaging, including hybrid devices that combine fluorescence with (i) hemodynamic or (ii) photoacoustic readouts, enabling simultaneous measurement of neural activity, blood flow, and oxygenation.

Figures were adapted from original publication with permission: a.i from Ref. [92], a.ii from Ref. [93], a.iii from Ref. [98], b from Ref. [38], c from Ref. [36], d.i from Ref. [106], d.ii from Ref. [74], d.iii from Ref. [107], e from Ref. [32], f.i from Ref. [108], f.ii from Ref. [109].

## Discussion

Head-mounted miniscopes have enabled numerous discoveries in neuroscience. The first device, developed in the early 2000s, was 7.5 cm long, weighted 25 g and suitable only for rats[7]. They have since evolved into compact and lightweight devices that can be fitted to the head of small animals such as mice and songbirds, with the imaging FOV, resolution and speed approaching those of the typical benchtop microscopes.

In this review, we introduce a unifying classification framework that organizes all currently published miniscope techniques into three major groups, based on their fundamental optical architectures. Within this framework, we highlight several key directions of technological development. The intent of this effort is twofold: for the biological community, it provides clarity within the diversity of device designs and offers a practical "search map" for selecting the right tool for specific research questions; for technology developers, it serves as a reference point to link new work to existing strategies, and perhaps to identify gaps or future opportunities for innovation.

From reviewing the past five years of advances, one conclusion from us is that the MEMS-based multiphoton point-scanning approach has rapidly emerged as one of the most intensively explored and promising technical roadmaps. This trend reflects its clear optical layout, the mature design theories, the availability of high-performance optical components, and the ease with which it can adapt advanced technologies developed in benchtop multiphoton microscopy. The recent commercialization of several MEMS-based miniscope systems indicates the transition from early prototyping to a stage of technological maturity. We therefore predict an acceleration of biological discoveries benefited by using MEMS-based 2P miniscopes in the next 3–5 years.

At the same time, there remains significant room for further technical improvement. Despite impressive progress in expanding spatial throughput, the number of neurons simultaneously recorded with current 2P miniscopes generally remains in the hundreds to around one-thousand, at least one order of magnitude below what state-of-the-art benchtop multiphoton systems (e.g., mesoscope[110,111], light-beads microscopy[80]) can achieve. This disparity constrains our ability to probe large-scale, coordinated population dynamics across complex networks. Similarly, while benchtop 2P methods have recently reached imaging rates of >1 kHz across tens to hundreds of neurons[78,81,112], thereby opening the door to capturing neuronal ensemble activity with near-electrophysiological precision, the temporal throughput of 2P miniscopes (product of number of recorded neurons and frame rate) is still one to two orders lower. With the rapid development of ultrafast neural activity indicators, including voltage[81,113-116] and glutamate sensors (e.g. iGluSnFR[117]), demand for high-speed miniscopes will only grow stronger. 1P miniscopes have benefited from their simple optics, reaching remarkable levels of miniaturization and weight reduction, making simultaneous multi-region recordings feasible. Whether 2P miniscopes can be pushed to similar levels of miniaturization while maintaining their high performance, remains a question for future exploration.

While miniscopes have made substantial advancements, with many new developments on the horizon, several challenges in their biology-level validation remain. One open question is whether the presence of these devices, despite being small and lightweight enough to minimize interference with movement, still affects animal behavior and neural activity. Addressing this issue calls for two complementary approaches. First, further miniaturization, such as wireless data transmission[13,51] and power delivery to eliminate the need for tethered cables, could reduce the device impact on the animal. Second, there is a need to develop standardized behavioral paradigms and quantitative metrics to assess the naturalness of animal behavior. Compared to the existing common practice of comparing the animal movement with and without the miniscope, such a standardization will enable a more rigorous evaluation of the impact of the miniscope, which in turn, will advance the frontier of the technology.

Another major challenge lies in assessing the signal fidelity of 1P miniscopes, particularly those relying on computational imaging, where signals can be heavily contaminated by background fluorescence. Imaging using both 2P and 1P modalities in the same FOV, preferably in a simultaneously manner, could serve as a valuable validation method for the signals extracted from 1P recordings[118-120].

Recently, researchers have begun to integrate single-cell–level optogenetics into miniscopes, aiming to simultaneously record and manipulate neuronal ensembles in freely behaving animals. However, the spatial precision of such approaches remains limited compared to benchtop systems. Moreover, rigorous biological validation to demonstrate that targeted stimulation could indeed produce selective and reproducible circuit-level effects in natural behaviors is still lacking. Addressing these challenges will be crucial before single-cell optogenetics with miniscopes can be widely adopted as a reliable tool in systems neuroscience.

Finally, to foster a robust ecosystem for the miniscope community, it would be beneficial to establish standardized interfaces and modular components, such as interchangeable objective lenses, fiber-to-miniscope connectors, detector modules, and animal interface systems. While there is a risk of over-standardization, such efforts could greatly accelerate the integration of new components into existing platforms. Additionally, the development of standardized protocols for animal surgeries and behavioral paradigms would further enhance reproducibility and collaboration across labs. Last but not least, the miniscope community have been, and will continue to be benefited from advances in fluorescent probes and imaging processing algorithms. Though not covered in this review, recent developments in image processing, particularly those leveraging deep learning approaches[119,121-123], have markedly improved background suppression, cell segmentation, and activity trace extraction. These innovations have substantially enhanced the fidelity of the extracted signals and the speed of data processing. With these multi-disciplinary effects and collaborations, miniscope technology is poised to further advance neuroscience research.

**Acknowledgement**

We acknowledge the support from The Research Council of Norway FRIPRO grant (ES737113, WZ), Proof of Concept grant (355770,WZ), Centre of Excellence grant (Centre for Algorithms in the Cortex 332640, WZ), National Infrastructure grant (NORBRAIN IV, WZ), National Institute of Neurological Disorders and Stroke (R01NS133924, R01NS118289, WY), National Eye Institute (R21EY029472, R01NS118289, WY), National Institute of Biomedical Imaging and Bioengineering (R21EB035306, WY), National Science Foundation (CAREER 1847141, WY), Burroughs Wellcome Fund (Career Award at the Scientific Interface 1015761, WY).

**Contributions**
W.Z and W.Y conceptualized the paper together; W.Y drafted the manuscript; W.Z prepared the figures; W.Z and W.Y discussed and edited the manuscript together and provided funding for the project.

**Table 1: State-of-the-art of 1P miniscope**

| | Large field of view | | | | | | | Multi-region | |
|---|---|---|---|---|---|---|---|---|---|
| Reference | Zhao Sci. Adv. 2025 [22] | Zhang Nature Bio. Med. Eng. 2024 [70] | Guo Sci. Adv. 2023 [18] | Zhang Nature Comm. 2023 [20] | Scherrer Nature Met. 2023 [19] | Rynes Nature Met. 2021 [17] | Scott Neuron 2018 [14] | Xue Natl. Sci. Rev. 2024 [21] | Hu Sci. Adv. 2025 [73] |
| Name | MiniXL | SOMM | Miniscope-LFOV | | Kiloscope | mini-mScope | cScope | TINIscope | mini-MCAM |
| Collection NA | 0.12 | ~0.1 | ~0.23* | 0.16 | Not available | Not available | 0.1 | Not available | Not available |
| Lateral resolution (µm) | 4.4 | 4 | 3.5 | 3 | ~4 | 39-56 | 14 | 2.19 | 10-20 |
| FOV ($mm^2$) | Φ 3.5 | 3.6x3.6 | 3.6x2.7 | Φ 3.6 | 4.8x3.6 | 8x10 | 7.8x4 | 0.45x0.45 | 4.5x2.55 |
| Frame rate (Hz) | 22 | 16 | 11 | NA. | NA. | Up to ~60 | 30-60 | 40 | 30 |
| Number of resolvable points (FOV/lateral resolution^2) [proportional to optical invariant] | $5.0\times10^5$ | $8.1\times10^5$ | $7.9\times10^5$ | $1.1\times10^6$ | $1.1\times10^6$ | $3.5\times10^4$ | $1.6\times10^5$ | $4.2\times10^4$ | $5.1\times10^4$ |
| Weight (g) | 3.5 | 2.5 | 13.9 | 0.5 | 1.4 | 3.8 | 33 | 0.43 | 1.175 |
| Magnification | 1.19x | 1x | 1.56x | 1x | 1x | <0.36x * | 0.7x | 2.4x | 0.4x * |
| Camera sensor | 5MP, 2.2 µm pixel size MT9P031 | 5MP, 2.2 µm pixel size MT9P031 | 5MP, 2.2 µm pixel size MT9P031 | 5 MP, 1.12 µm pixel size GC5035 | 0.48 MP, 4.8 µm pixel size NOIP1SP0480A | 0.36 MP, 6 µm pixel size MT9V032 | 0.36 MP, 6 µm pixel size | 0.31 MP, 3 µm pixel size OV7251 | 0.92 MP, 1.4 µm pixel size KD-B209M3-76 camera module |
| Note | ±200 µm adjustment of focal depth. | Extended DOF over 300 µm. | For rats. ±100 µm adjustment of focal depth. | Extended DOF over 300 µm. | | | For rats. | 4 TINIscopes used together for multi-region imaging | 4 modules used together for multi-region imaging |

* Estimated value based on the provided information in the literature.

## Table 2: State-of-the-art of 2p/3p miniscope

| | Wu Nature Met. 2025 [36] | Madruga Nature Comm 2025 [35] | Zong Cell 2022 [34] | Zhao Nature Met. 2023 [39] | Klioutchnikov Nature Met. 2023 [38] | Li Optica 2021 [32] | Blot preprint 2025 [74] | Ozbay Sci. Rep. 2018 [30] |
|---|---|---|---|---|---|---|---|---|
| | FHIRM-TPM 3.0 | UCLA 2P Miniscope | MINI2P | m3PM | | 2P fiberscope | 2P-FENDO-II | 2P-FCM |
| Modality | 2P | 2P | 2P | 3P | 3P | 2P | 2P | 2P |
| Scanning | MEMS | MEMS | MEMS | MEMS | MEMS | Fiber scanning | Benchtop relayed through multi-core fiber | Benchtop relayed through multi-core fiber |
| Excitation NA | 0.7/0.5/0.26 | 0.36 | 0.36 | 0.55 | 0.48 | 0.51/ 0.33 | Not available | 0.45 |
| Collection NA | 0.74/0.67 /0.4 | 0.6 | 0.5 | 0.65 | 0.9 | 0.8/0.5 | Not available | 0.45 |
| Lateral resolution (µm) | 0.68/0.79 /1.46 | 0.98 | 1.15-1.24 | 0.97 | 1.18 | 0.9/1.1 | 3.9 | 1.8 |
| Axial resolution (µm) | 3.73/7.04 /23.68 | 10.18 | 12.8-17.8 | 7.21 | 13.8 | 5/14.4 | 9-13 | 10 |
| FOV ($mm^2$ for rectangle or mm for diameter) | 0.3x0.26/ 0.5x0.425/ 1x0.8 | 0.445x0.38 | 0.5x0.5 / 0.42x0.42 | 0.4x0.4 | 0.3x0.3 | Φ 0.14 / Φ 0.3 | Φ 0.48 | Φ 0.24 |
| Axial range (mm) | NA | 0.15 | 0.24 | None | 0.6 | None | None | 0.18 |
| Frame rate (Hz) | 20 @ 256 lines | 8.62 @ 512x354 | 15 /40 @ 256x256 | 15.93 @ 128x128 | 10.6 @ 273x280 | 2 / 3 | 10 | 2.5 |
| Number of resolvable points (FOV/lateral resolution ^2) [proportional to optical invariant] | $1.69\times10^5$/ $3.40\times10^5$/ $3.75\times10^5$ | $1.76\times10^5$ | $1.75\times10^5$ | $1.70\times10^5$ | $6.45\times10^4$ | $5.84\times10^4$ | $1.19\times10^4$ | $1.40\times10^4$ |
| Weight (g) | 2.6 | 4 | 2.4 | 2.17 | 2 | 0.6 | 1.4 | 2.5 |

**Figures and Figure caption**

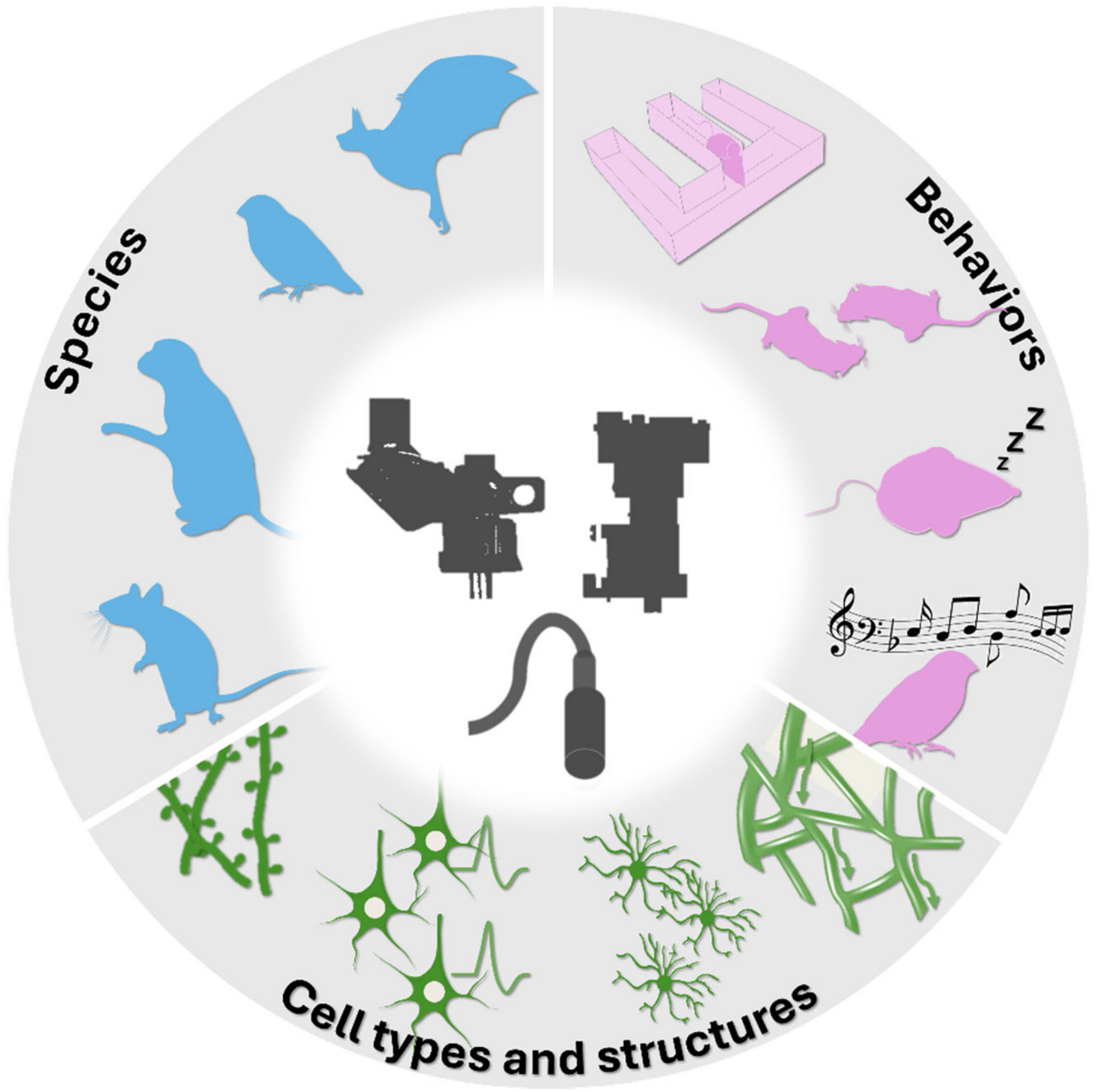

**Figure 1 | Applications of head-mounted miniscopes in neuroscience.**
Miniscopes enable recording of neural activity across diverse species (mice, rats, songbirds, bats, and non-human primates) during behaviors that are difficult or impossible to study in head-fixed conditions, such as spatial navigation, social interaction, sleep, and vocal communication. Combined with genetically encoded or vascular indicators, they allow imaging of neurons, astrocytes, microglia, dendrites, spines, and blood vessels. This versatility has made it possible to study neural dynamics at multiple scales, from subcellular compartments to local circuits and distributed brain networks in freely behaving animals.

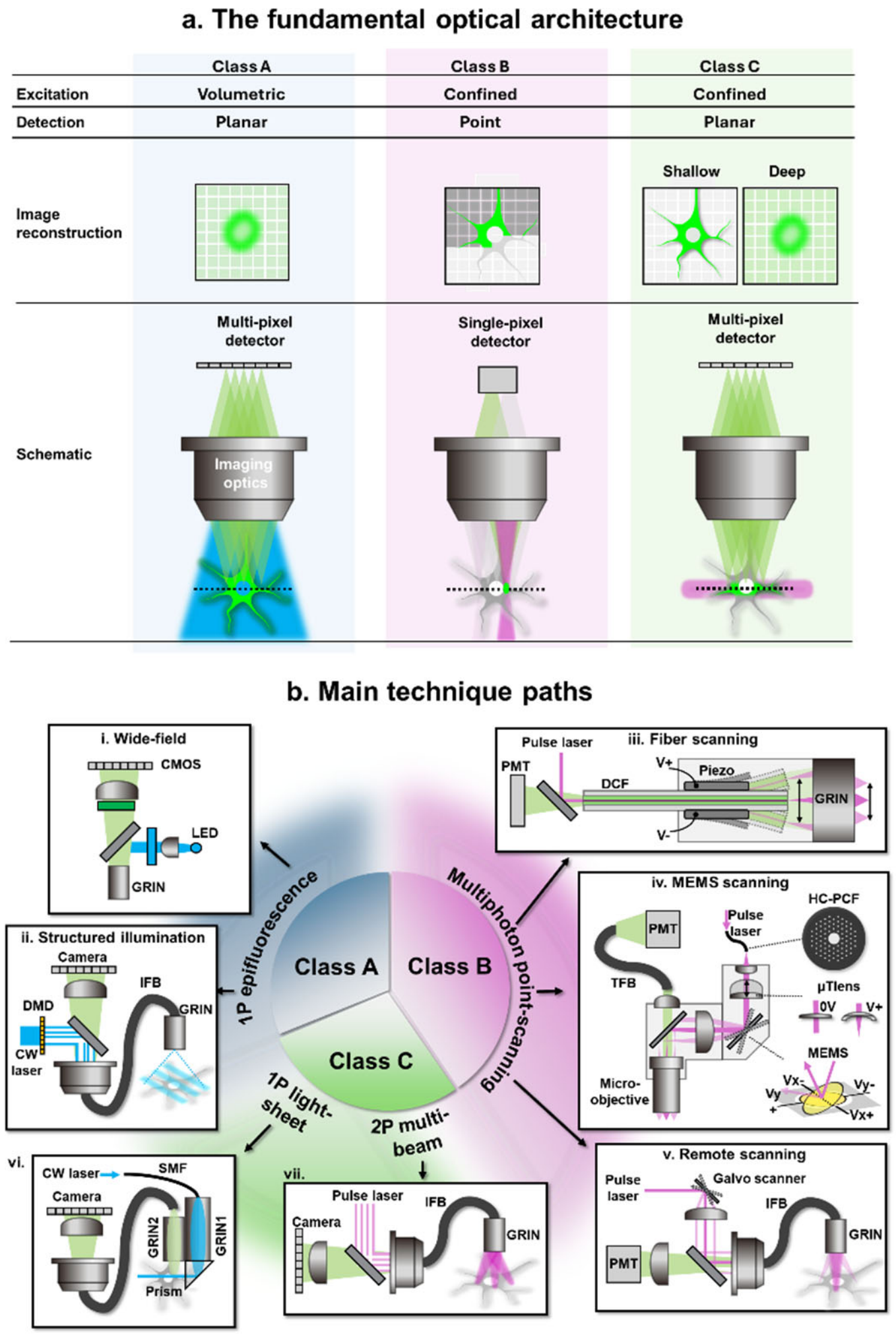

**Figure 2 | Core architecture and representative designs of miniscopes.**

**(a)** Conceptual classification of miniscope architectures. Class A uses wide-field illumination with planar detection (e.g., 1P epifluorescence miniscopes), offering fast imaging but with strong neuropil and out-of-focus contamination. Class B employs point excitation and single-pixel detection (e.g., 2P/3P point scanning miniscopes), providing high resolution and depth penetration at the cost of slower imaging speed. Class C combines confined excitation with planar detection (e.g., light-sheet or fiber-bundle–based systems), balancing speed and resolution in shallow tissue but with degraded image quality at depth.

**(b)** Representative optical layouts. Class A includes (**i**) basic 1P wide-field miniscopes and (**ii**) structured illumination designs using digital micromirror devices (DMDs) and imaging fiber bundles (IFBs). Class B comprises multiphoton devices with different scanning strategies: (**iii**) fiber scanning using a piezo-actuated, (**iv**) MEMS (miniaturized resonant mirrors) scanning, and (**v**) remote scanning (beam steered by bench-mounted scanners and relayed to the sample by IFBs). With MEMS-scanning strategy, the state-of-art systems applied hollow-core photonic crystal fibers (HC-PCF) to deliver femtosecond lasers and electrically tunable lenses (uTLens) to adjust imaging depth. Class C includes (**vi**) 1P light-sheet systems and (**vii**) multi-beam remote-scanning 2P variants that spread excitation across multiple fiber cores for camera-based detection.

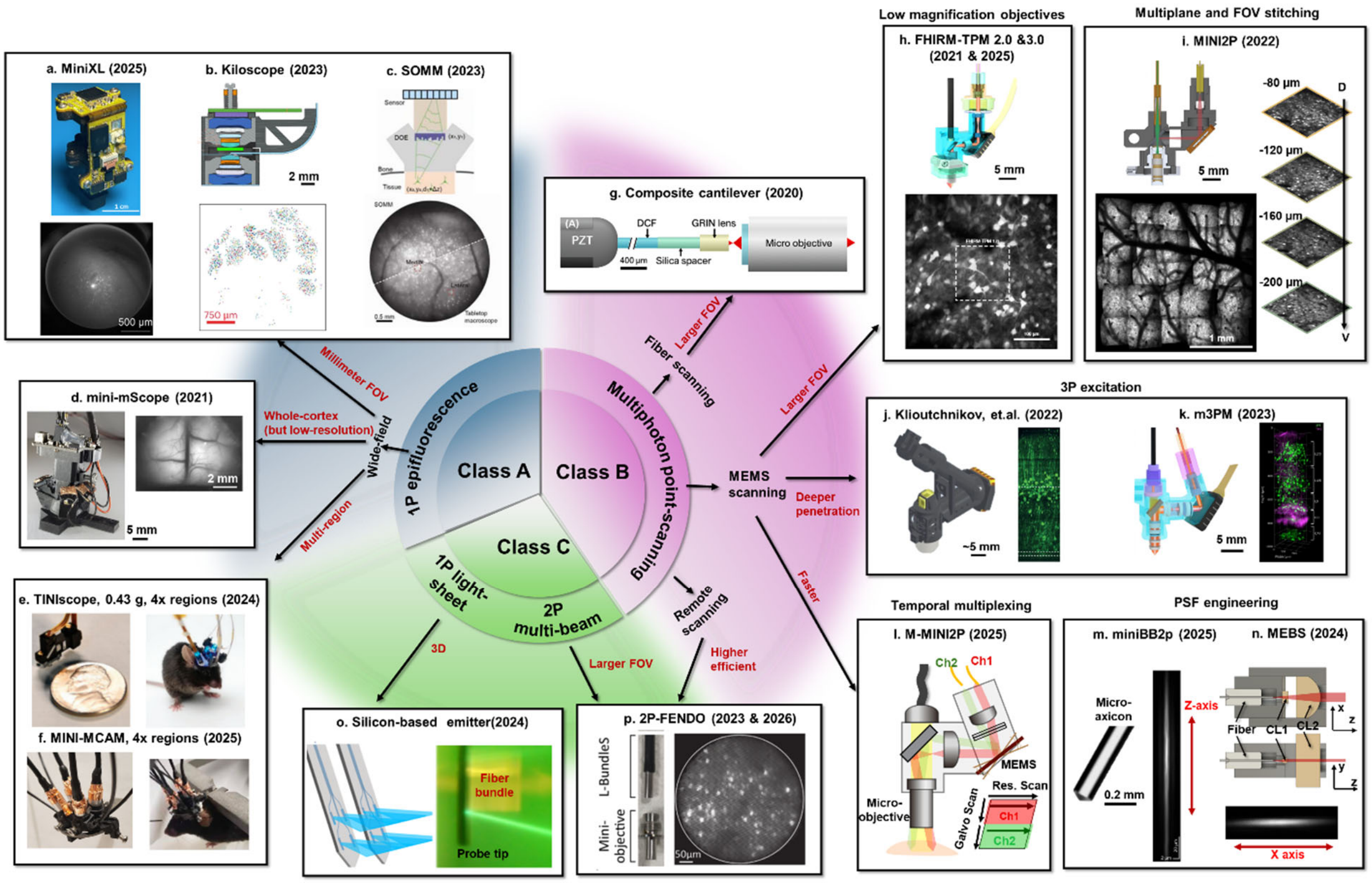

**Figure 3 | Featured development by engineering better optical components in the last 5 years.**

(**a-c**) In Class A, recent devices extend the field of view (FOV) to the mesoscale (a few millimeters), while maintaining single-cell resolution in sparsely labeled samples. (**d**)The brain-wide (>10mm) FOV with lower resolution (>10s µm) has also been achieved. (**e** and **f**) Moreover, ultralight miniscope designs have enabled (**e**) 4-region (TINIscope) and (**f**) 4-region mini-MCAM simultaneously recording. (**g**) For increasing the FOV within fiber scanning category in class B, a composite cantilever was developed to engineer the fiber tip by attaching a beam expander and focusing optics to it. (**h-n**) In class B, the development of the MEMS-based 2P systems appears in three major directions, a continue increasing of FOV using (**h**) low magnification objectives and (**i**) applying multiplane imaging and FOV stitching, an increase of penetration depth by three-photon excitation (**j** and **k**), and an increase in imaging speed by (**l**) beam multiplexing and (**m** and **n**) PSF-engineering. In Class C, new developments include (**o**) thin photonic probes with silicon grating-based light-sheet delivery and light field detection to achieve 3D imaging. There is also a new series of fiber-bundle–based multiphoton systems (**p**) with increased FOV and efficiency, classified between Class B and C because it uses fiber-bundle-camera-based planar detection relays with multi-beam multiphoton excitation. Figures were adapted from original publication with permission: a from Ref. [22], b from Ref. [19], c from Ref. [70], d from Ref. [17], e from Ref. [21], f from Ref. [73], g from Ref. [31], h from Ref. [33], i from Ref. [34], j from Ref. [38], k from Ref. [39], l from Ref. [55], m from Ref. [88], n from Ref. [89], o from Ref. [76], and p from Ref. [74].

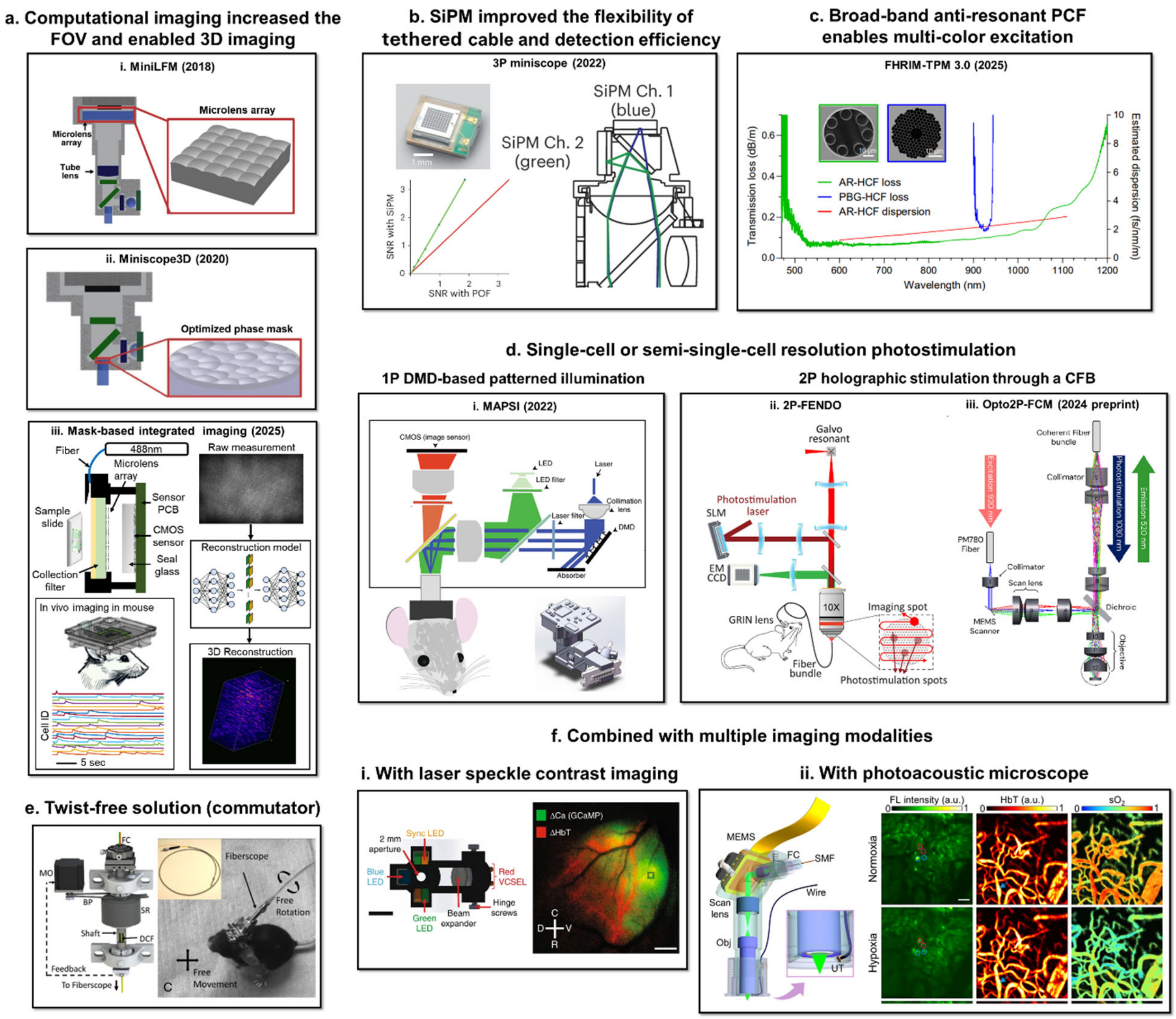

**Figure 4 | Emerging technologies expanding the capabilities of head-mounted miniscopes.**

- **(g)** Computational imaging approaches that rely on engineered point spread functions and computational reconstruction.
- **(h)** Integration of on-board detectors such as silicon photomultipliers (SiPMs) improves fluorescence detection efficiency, eliminates bulky collection fibers, and enables multichannel recording.
- **(i)** Advances in hollow-core fibers (HCF), including anti-resonant HCF (AR-HCF), increased transmission bandwidth, allowing multicolor multiphoton excitation. PBG-HCF, photonic bandgap photonic crystal fiber (same concept as HC-PCF).
- **(j)** Integration of (**i**) 1P and (**ii** and **iii**) 2P patterned optogenetics in miniscopes enables simultaneous recording and targeted manipulation of neuronal ensembles in freely behaving animals. SLM, spatial light modulator.
- **(k)** Engineering of optical-electronic hybrid commutators supports twist-free animal movement with minimal loss of optical signals, facilitating long-duration behavioral studies. FC, fiber collimator; MO, motor; BP, belt and pulley; SR, slip ring, and DCF, double-clad fiber.

**(l)** Multimodal miniscopes expand beyond calcium imaging, including hybrid devices that combine fluorescence with (i) hemodynamic or (ii) photoacoustic readouts, enabling simultaneous measurement of neural activity, blood flow, and oxygenation.
Figures were adapted from original publication with permission: a.i from Ref. [92], a.ii from Ref. [93], a.iii from Ref. [98], b from Ref. [38], c from Ref. [36], d.i from Ref. [106], d.ii from Ref. [74], d.iii from Ref. [107], e from Ref. [32], f.i from Ref. [108], f.ii from Ref. [109].